\documentstyle[10pt,bezier,epsfig]{article}
\begin{document}
\def \inbar{\vrule height1.5ex width.4pt depth0pt}
\def \xC{\relax\hbox{\kern.25em$\inbar\kern-.3em{\rm C}$}}
\def \xR{\relax{\rm I\kern-.18em R}}
\newcommand{\xZ}{Z \hspace{-.08in}Z}
\newcommand{\xbe}{\begin{equation}}
\newcommand{\xee}{\end{equation}}
\newcommand{\xbea}{\begin{eqnarray}}
\newcommand{\xeea}{\end{eqnarray}}
\newcommand{\xnn}{\nonumber}
\newcommand{\xkt}{\rangle}
\newcommand{\xbr}{\langle}
\newcommand{\xlll}{\left( }
\newcommand{\xrrr}{\right)}
\newcommand{\xcun}{\mbox{\footnotesize${\cal N}$}}
\title{Inverting  Time-Dependent Harmonic Oscillator Potential
by a Unitary Transformation and a New Class of Exactly~Solvable~Oscillators}
\author{Ali Mostafazadeh\thanks{E-mail: alimos@phys.ualberta.ca}\\ \\
Theoretical Physics Institute, University of Alberta, \\
Edmonton, Alberta,  Canada T6G 2J1.}
\date{November 1996}
\maketitle

\begin{abstract}
A time-dependent unitary (canonical) transformation is found which
maps the Hamiltonian for a harmonic oscillator with  time-dependent
real mass and real frequency to that of a generalized harmonic
oscillator with time-dependent real mass and imaginary frequency.
The latter may be reduced to an ordinary harmonic oscillator by
means of another unitary (canonical) transformation. A simple
analysis of the resulting system leads to the identification of a
previously unknown class of exactly solvable time-dependent
oscillators. Furthermore, it is shown how one can apply these
results to establish a canonical equivalence between
some real and imaginary frequency oscillators. In particular  it is
shown that a harmonic oscillator whose frequency is constant
and whose mass grows linearly in time is canonically equivalent
with an oscillator whose frequency changes from being real to
imaginary and vice versa repeatedly.
\end{abstract}



The solution of the Schr\"odinger equation, $H\psi=i\dot\psi$, for a
harmonic oscillator  with  time-dependent mass $m$ and frequency
$\omega$, i.e.,
	\xbe
	H(t)=\frac{1}{2m(t)}\,p^2+\frac{m(t)\omega^2(t)}{2}\,x^2\;,
	\label{H}
	\xee
has been the subject of continuous investigation since late 1940's,
\cite{old,l-r,kim,j-k}. The main reason for the interest in this problem is its wide
range of application in the description of physical systems. 
Although by now there exist dozens of articles on the subject, a
closed analytic expression for the time-evolution operator is still missing.
Recently, Ji and Kim \cite{j-k} showed that using the Lewis-Riesenfeld
method \cite{l-r} one can construct an invariant operator in terms of the
(two independent) solutions of the classical dynamical equations:
	\xbe
	\frac{d}{dt}[m(t)\,\frac{d}{dt}\, x_c(t)]+m(t)\omega^2(t)x_c(t)=0\;,
	\label{class}
	\xee
and therefore reduce the solution of the  Schr\"odinger equation
to that of Eq.~(\ref{class}), for $\omega(t),m(t)\in\xR$.  The case where
the frequency $\omega$ is imaginary has been considered only in
the time-independent case \cite{imaginary}.

The purpose of this note is to study the implications of  the recently
developed method of adiabatic unitary transformation of the Hilbert
space \cite{p16} for this problem. The basic idea of this method is to
use the inverse of the adiabatically approximate time-evolution
operator to transform to a moving frame. This transformation has
proven to lead to some interesting results for the system consisting
of a magnetic dipole in a changing magnetic field. The analogy
between the dipole system and the time-dependent harmonic oscillator
is best described in terms of the relation between their dynamical
groups, namely $SU(2)$ and $SU(1,1)$, \cite{jackiw}.

Let us first concentrate on the real frequency case, $\omega(t)\in\xR$. 
Then the Hamiltonian (\ref{H}) and its eigenvectors $|n;t\xkt$ can be
expressed in terms of the creation ($a^\dagger$) and annihilation
($a$) operators, namely
	\xbea
	H(t)&=&\omega(t)[a^\dagger(t)a(t)+1/2]\;,
	\label{H1}\\
	|n;t\xkt&=&\frac{1}{\sqrt{n!}}\, a^{\dagger n}(t)|0;t\xkt\;,
	\label{eg-ve}
	\xeea
where
	\xbe		
	a(t):=\frac{1}{\sqrt{2}}\,(e^{\kappa(t)} x+ie^{-\kappa(t)}p)\;,~~~~~
	\kappa(t):=\frac{1}{2}\ln [m(t)\omega(t)]\;,
	\label{a}
	\xee
and $|0;t\xkt$ corresponds to the ground state at time $t$, i.e., $a(t)|0;t\xkt=0$.
Similarly to the time-independent case the eigenvalue equation
	\xbe
	H(t)|n;t\xkt=E_n(t)|n;t\xkt\;,
	\label{eg-va-eq}
	\xee
is satisfied for $E_n(t)=E_n[\omega(t)]=\omega(t)(n+1/2)$. Therefore
the eigenvalues are non-degenerate and the adiabatically approximate
time-evolution operator \cite{p16} may be expressed as:
	\xbe
	U_0(t):=\sum_ne^{i\alpha_n(t)}|n;t\xkt\xbr n;0|\;,
	\label{u0}
	\xee
where 
	\xbea
	\alpha_n(t)\::=\:\delta_n(t)+\gamma_n(t)\;,~~~~
	\delta_n(t)&:=&-\int_0^t E_n(t')dt'\,~~~~
	\gamma_n(t)\::=\:i\int_0^t A_{nn}(t')dt'\,,
	\label{adg}\\
	A_{mn}(t)&:=&\xbr m;t|\frac{d}{dt}|	n;t\xkt\;.
	\label{amn}
	\xeea

Following the ideas developed in Ref.~\cite{p16}, let us next 
use $U_0^{-1}(t)=U_0^{\dagger}(t)$ to perform a unitary
transformation of the Hilbert space.  Recall that under a
general unitary transformation $|\psi(t)\xkt\to|\psi'(t)\xkt=
{\cal U}(t)|\psi(t)\xkt$, the Hamiltonian $H$ and the
corresponding time-evolution operator transform according to
	\xbea
	H(t)\rightarrow H'(t)&=&
	{\cal U}(t)H(t)\,{\cal U}^\dagger(t)-i\,{\cal U}(t)\,\dot{\cal U}^\dagger(t)\;,
	\label{trans}\\
	U(t)\rightarrow U'(t)&=&{\cal U}(t)U(t)\,{\cal U}^\dagger(0)\;.
	\label{trans-u}
	\xeea
Substituting $U_0^\dagger$ for ${\cal U}$ in (\ref{trans}), one can
show, \cite{p16}, that
	\xbe
	H'(t)=-i\sum_{n\neq m}e^{-i[\alpha_m(t)-\alpha_n(t)]}A_{mn}(t)
	|m;0\xkt\xbr n;0|\;.
	\label{trans-H}
	\xee

In order to express the transformed Hamiltonian $H'$ in a closed
form, one must first compute $A_{mn}$. This can be done directly
by substituting Eq.~(\ref{eg-ve}) in (\ref{amn}) or indirectly using
the identity
	\[A_{mn}(t)=\frac{\xbr m;t|\frac{d H(t)}{dt}|n:t\xkt}{E_n(t)-E_m(t)}\;,
	~~~~~{\rm for}~~~m\neq n\;,\]
which is obtained by differentiating both sides of Eq.~(\ref{eg-va-eq})
and making use of the orthonormality of  $|n;t\xkt$. The latter method
turns out to be much simpler. It yields
	\xbe
	A_{mn}(t)=\frac{\dot\kappa(t)}{2}\left[\sqrt{n(n-1)}\delta_{m, n-2}-
	\sqrt{m(m-1)}\delta_{m-2,n}\right]\,.
	\label{amn=}
	\xee
Here use is made of the well-known relations
	\xbe
	a(t)|n;t\xkt=\sqrt{n}\:|n-1;t\xkt\;,~~~
	 a^\dagger(t)|n;t\xkt=\sqrt{n+1}\:|n+1;t\xkt\;,~~~{\rm and}~~~
	\dot  a(t)=\dot\kappa(t) a^\dagger(t)\;.
	\label{rel}
	\xee
Note that  Eq.~(\ref{amn=}) is also valid for $m=n$ since due to
the fact that $|n;t\xkt$ can be chosen to be real, $A_{nn}(t)=0$.
In particular $\gamma_n(t)=0$. Hence, $\alpha_n(t)=
\delta_n(t)=(n+1/2)\delta(t)$, where
	\xbe
	\delta(t):=-\int_0^t\omega(\tau)d\tau\;.
	\label{delta}
	\xee

Substituting the expressions for $\alpha_n$ and $A_{mn}$ in
Eq.~(\ref{trans-H}) and performing the summation over $m$, one
finds
	\xbea
	H'(t)&=&\frac{-i\dot\kappa(t)}{2}\,\sum_n \left[
	\sqrt{n(n-1)}e^{2i\delta(t)}|n-2;0\xkt\xbr n;0|-
	\sqrt{(n+1)(n+2)}e^{-2i\delta(t)}|n+2;0\xkt\xbr n;0|\right]\;,\xnn\\
	&=&\frac{-i\dot\kappa(t)}{2}\,\left[
	 e^{2i\delta(t)} a^2(0)-e^{-2i\delta(t)}a^{\dagger 2}(0)\right]\;,\xnn\\
	&=&\frac{\dot\kappa(t)}{2}\left\{\sin[2\delta(t)](
	e^{-2\kappa(0)}p^2-e^{2\kappa(0)}x^2)+\cos[2\delta(t)]
	(xp+px)\right\}\;,
	\label{H'=}
	\xeea
Next consider the Schr\"odinger equation in the transformed
frame: $H'(t)|\psi'(t)\xkt=i|\dot\psi'(t)\xkt$. The presence of a
total derivative $\dot\kappa$ on the right hand side of Eq.~(\ref{H'=}),
suggests a redefinition of the time  $t\to t':=\kappa(t)$. Note
that for  $\dot\kappa(t)=0$, i.e.,  $\omega(t)=c/m(t)$ for some
constant $c\in\xR$, $H'(t)$ vanishes identically and the adiabatic
approximation is exact. Furthermore, for $\dot\kappa<0$, one can
consider the time-reversed system for which $\dot\kappa>0$.
The time-evolution operator for the original system is obtained
from that of the time-reversed system by inversion. Therefore,
without loss of generality, one can assume $\dot\kappa(t)>0$.
The latter allows for the above-mentioned redefinition of time
$t\to t'$. This leads to the Schr\"odinger equation
$h(t')|\psi'(t')\xkt=i|\dot\psi'(t')\xkt$ for the Hamiltonian
	\xbea
	h(t')&:=&\frac{1}{2}\left\{\sin[2\tilde\delta(t')](
	e^{-2\kappa_0}p^2-e^{2\kappa_0}x^2)+\cos[2\tilde\delta(t')]
	(xp+px)\right\}\;,\xnn\\
	&=&\frac{1}{2m'(t')}\,p^2+\frac{m'(t')\omega^{'2}(t')}{2}\,x^2+
	\frac{1}{2}\,\beta(t')(xp+px)\;,
	\label{tilde-H'}
	\xeea
where
	\xbea
	\tilde\delta(t')&:=&\delta(t(t'))\,,~~~~\kappa_0\::=\:\kappa(0)\;,
	~~~~m'(t')\::=\:\frac{e^{2\kappa_0}}{\sin[2\tilde\delta(t')]}\;,
	\label{m'=}\\
	\omega'(t')&:=&i\sin[2\tilde\delta(t')]\,,~~~~
	\beta(t')\::=\:\cos[2\tilde\delta(t')]\:=\:\sqrt{1+\omega^{'2}(t')}\;.
	\xeea
Hence, the transformed Hamiltonian is a generalized
harmonic oscillator \cite{jackiw}:
	\xbea
	h(t')&=&\frac{1}{2}\:\left[\alpha(t')p^2+\beta(t')(xp+px)+
	\gamma(t')x^2\right]\;,
	\label{gho}\\
	\alpha(t')&:=&1/m'(t')\;,~~~~~\gamma(t')\::=\:m'(t')\omega^{'2}(t')
	\xnn
	\xeea
with a real mass $m'(t')$ and an imaginary frequency $\omega'(t')=i
e^{2\kappa_0}/m'(t')$. Note that as a result of the adiabatic unitary
transformation and redefinition of time, the two arbitrary functions
$m(t)$ and $\omega(t)$ have been reduced to a single function namely
$\tilde\delta(t')$. 

It is well-known that one can transform the generalized harmonic
oscillator  to an ordinary harmonic oscillator by the time-dependent
canonical transformation, \cite{jackiw},
	\xbe
	x\to x\;,~~~~~~~ p\to p+\left[ \frac{\beta(t')}{\alpha(t')}\right]\,x\;.
	\label{can}
	\xee
This leads to the Hamiltonian
	\xbe
	h'(t')=\frac{1}{2}\:\left\{ \alpha(t')p^2+\left[\gamma(t')-
	\frac{\beta^{2}(t')}{\alpha(t')}-\frac{d}{dt}\left(
	\frac{\beta(t')}{\alpha(t')}\right)
	\right]x^2\right\}=\frac{1}{2m'(t')}\,p^2+\frac{m'(t')\Omega^{'2}(t')}{2}x^2\;,
	\label{h'}
	\xee
where
	\xbe
	\Omega'(t'):=\sqrt{-1+\frac{2}{\sin[2\tilde\delta(t')]}
	\frac{d\tilde\delta(t')}{dt'}}\;.
	\label{Omega}
	\xee
Note that $\Omega'$ can be real or imaginary depending on the form
of $\tilde\delta(t')$. Clearly for $\Omega'=0$, the problem reduces to
a free particle with a variable mass whose solution can be exactly
given. In terms of the original functions, the condition $\Omega'=0$
is expressed as
	\xbe
	m(t)=\left[\frac{m_0}{\omega(t)}\right]\tan^2\int_0^t\omega(\tau)d\tau\;,
	\label{condi}
	\xee
where $m_0$ is a real constant. 

Eq.~(\ref{condi}) determines a new class of exactly solvable cases
which is the analog of the exactly solvable magnetic dipole Hamiltonians
obtained in \cite{p16}. The only difference is that here an additional
canonical transformation (\ref{can}) is also performed.  It is not difficult to
see that this transformation corresponds to the action of the
unitary operator ${\cal U}'(t'):=\exp[i\beta(t')x^2/(2\alpha(t))]$ on the Hilbert
space. In fact, the Hamiltonian $h'$ can be obtained from $h$ by
substituting $h$ for $H$ and ${\cal U}'$ for ${\cal U}$ in Eq.~(\ref{trans}).
Similarly in view of Eq.~(\ref{trans-u}), one has the following relation
between the evolution operators
	\xbe
	u(t')=e^{-i\frac{\beta(t')}{2\alpha(t')}\,x^2}u'(t')\:
	e^{i\frac{\beta(\kappa_0)}{2\alpha(\kappa_0)}\,x^2}\;,
	\label{u=u'}
	\xee
where $u(t')$ and $u'(t')$ are the evolution operators for the
Hamiltonians $h(t')$ and $h('(t')$, respectively. A further
application of  Eq.~(\ref{trans-u}) leads to the expression
	\xbe
	U(t)=U_0(t)u(t'(t))\;,
	\label{u=uu}
	\xee
for the time-evolution operator associated with the original
Hamiltonian (\ref{H}). Here $U_0(t)$ is the adiabatically approximate
expression for the time-evolution operator given by Eq.~(\ref{u0}). 

For the cases where the condition (\ref{condi}) holds, $u'(t')$
is the evolution operator for a free particle with a variable mass
$m'(t')$, i.e.,
	\xbe
	u'(t')=\exp\left[\frac{-i}{2}\int_{\kappa_0}^{t'}
	\frac{d\tau}{m'(\tau)}~p^2\right]\;.
	\label{u'=p2}
	\xee

If $\Omega'$ does not vanish but satisfies $\Omega'(t')=
\Omega'_0/m'(t')$ for some real constant $\Omega'_0$, then the
Schr\"odinger equation for $h'$ is still exactly solvable. In fact in this
case
	\[ u'(t')=\exp\left[\frac{-i}{2}\int_{\kappa_0}^{t'}
	\frac{d\tau}{m'(\tau)}~(p^2+\Omega^{'2}_0x^2)\right]\;.\]
The condition $\Omega'(t')=\Omega'_0/m'(t')$ is a generalization
of $\Omega'(t')=0$. It corresponds to a larger class of exactly solvable
time-dependent harmonic oscillators. In terms of the original
parameters $m$ and $\omega$ this condition is expressed as:
	\xbe
	m(t)=\frac{m_0\:f^{\frac{1}{\sqrt{\zeta-1}}}(t)}{
\omega(t)g^{\frac{1}{\sqrt{\zeta}}}(t)}\:,
	\label{condi-g}
	\xee
where $m_0$ and $\zeta:=1+e^{4\kappa_0}/\Omega_0^2$ are
positive constants  and
	\[ 
	f(t):=\frac{\sqrt{\zeta-z(t)}-\sqrt{\zeta-1}}{\sqrt{\zeta-z(t)}+
	\sqrt{\zeta-1}}\;,~~~
	g(t):=\frac{\sqrt{\zeta-z(t)}-\sqrt{\zeta}}{\sqrt{\zeta-z(t)}+
	\sqrt{\zeta}}\;,~~~
	z(t):=1-\frac{1}{\zeta-1}\:\sin^2\int_0^t\omega(\tau)d\tau\;.\]
Eq.~(\ref{condi}) is a special case of (\ref{condi-g}) where
$\Omega'_0\to 0$.

Finally I would like to emphasize the following points:
	\begin{itemize}
	\item[1)] For the case that $\Omega'$ is real, one can repeat 
the above analysis by replacing $H$ of Eq.~(\ref{H}) by $h'$ of
Eq.~(\ref{h'}).  In principle this may lead to yet other exactly
solvable cases. If the iteration of this procedure yields oscillators
with real frequency at each step, then it can be repeated
indefinitely.  This leads to a product expansion for the
time-evolution operator which is analogous to what is called an
adiabatic product expansion in Ref.~\cite{p16}. 
	\item[2)] The above analysis also indicates that the
time-dependent harmonic oscillators whose mass $M$ and frequency
$\Omega$ are related according to
	\xbe
	\Omega=\sqrt{-1-\frac{\dot M}{\sqrt{M^2-M_0^2}}}\;,	\label{Condi}
	\xee
play a universal role. This is because as shown above the problem
for the most general real frequency harmonic oscillator  (\ref{H}) can
be reduced to this case by means of a series of unitary (canonical)
transformations. Eq.~(\ref{Condi}) is obtained from (\ref{Omega}) by
expressing the right hand side of (\ref{Omega}) in terms of  the mass.
The parameter $M_0$ is an arbitrary positive constant corresponding
to $e^{2\kappa_0}$.
	\item[3)] It is tempting to seek applications of the known results for
the real frequency oscillators to the time-dependent imaginary frequency
oscillators satisfying (\ref{Condi}). In view of the above analysis, there
is a class of imaginary frequency oscillators of this form which 
are canonically equivalent to some real frequency oscillators. 
One must however be aware that by performing the canonical
transformation described in this article in the reverse order, one might
not be able to transform an arbitrary imaginary frequency oscillator to
a real frequency one, even if it satisfies Eq.~(\ref{Condi}). In this case
the transformation associated with the adiabatic
approximation would not  be the same as the one obtained above.
Nevertheless, the above scheme is consistent in the sense that
if the frequency of the oscillator obtained by transforming back an imaginary
frequency oscillator satisfying (\ref{Condi}) turns out to be real then one
has the desired result. Otherwise, the method fails to transform the
imaginary frequency oscillator to a real frequency one. A simple example of
a case where Eq.~(\ref{Condi}) is satisfied but the transformation to a real
frequency oscillator is not possible is the time-independent oscillator with
$M=$ const.\ and $\Omega=i$. Another example is  the case where the
mass is decaying in time $t'$ according to $M=m'(t')=M_0(1+e^{-\mu t'})$
with $\mu>1$ and the frequency $\Omega=\Omega'(t')$ is given by
Eq.~(\ref{Condi}),  i.e., $\Omega^2 =-1+\mu(1+2e^{\mu t'})^{-1/2}$. This
system is particularly interesting since the potential $V:=M\Omega^2/2$
changes sign at $t'=\kappa_*:=\frac{1}{\mu}\ln(\frac{\gamma^2-1}{2})$.
It is positive for $t'<\kappa_*$ and negative for $t'>\kappa_*$. Hence,
at $t'=\kappa_*$ the energy spectrum undergoes a `phase transition.'
However, it is not difficult to see that transforming this system  to a real
frequency oscillator is not possible. This is because making the
necessary canonical transformations, one obtains an oscillator with the
mass $m(t)$ and frequency $\omega(t)$ satisfying:
	\xbe
	[m(t)\omega(t)]^{-2/\mu}=-1-\frac{1}{\sin 2\int_0^t
	\omega(\tau)d\tau}.
	\label{condi-3}
	\xee
Thus one should seek functions $m:[0,\infty)\to\xR^+$ and
$\omega:[0,\infty)\to\xR^+$ which satisfy Eq.~(\ref{condi-3})
and are positive for $t\in[0,T]$ for some $T>t_*$, where $t_*$
is defined according to $\kappa(t_*)=\kappa_*$. That such
functions do not exist can be directly inferred from the fact that
for $t\to 0$ the right hand side of Eq.~(\ref{condi-3}) tends to
$-\infty$, whereas the left hand side remains positive. A simple
example of a real frequency oscillator which is canonically equivalent
to an oscillator whose frequency fluctuates between real and imaginary
values is 
	\xbe
	m(t)=m_0+\mu\, t\;,~~~~\omega(t)=\omega_0\;,
	\label{osc}
	\xee
where $m_0,~\mu$, and $\omega_0$ are positive constants.
Applying the canonical transformations introduced in this
article to this oscillator, one arrives at another canonically
equivalent oscillator with mass $m'(t')$ and frequency
$\Omega'(t')$ given by
	\xbe
	m'(t')=\frac{m_0\omega_0}{\sin\left[ \frac{2}{\mu}(m_0
	\omega_0-e^{2t'})\right]}\;,~~~~
	\Omega^{'2}(t')=-1-\frac{2e^{2t'}}{\mu \sin\left[ \frac{2}{\mu}(m_0
	\omega_0-e^{2t'})\right]}\;,
	\label{m-o}
	\xee
where $t'=\kappa(t)=\ln[\omega_0(m_0+\mu t)]/2$, and use
is made of Eqs.~(\ref{m'=}) and (\ref{Omega}). Fig.~1 shows a
plot of $\Omega^{'2}$ as a function of $t'$, for $m_0=\mu=
\omega_0=1$. As seen from this plot, $\Omega^{'2}$ changes sign
repeatedly. However, the original oscillator  (\ref{osc}) is a very simple
time-dependent oscillator with positive real mass and frequency.
In fact, one might try to apply the results of  Refs.~\cite{kim,j-k} to obtain
exact solution of the Schr\"odinger equation for this oscillator.
This would immediately lead to the exact solution of the Schr\"odinger
equation for the oscillator (\ref{m-o}), at least for the periods of time during
which $m'$ and $\Omega'$ are continuous functions of $t'$. This is
particularly interesting since as $\Omega^{'2}$ changes sign from
positive to negative, the spectrum of the corresponding oscillator
(\ref{m-o}) changes from being discrete to continuous and vice versa,
while the spectrum of the canonically equivalent oscillator
(\ref{osc}) remains always discrete.
	\end{itemize}
	\begin{figure}
	\epsffile{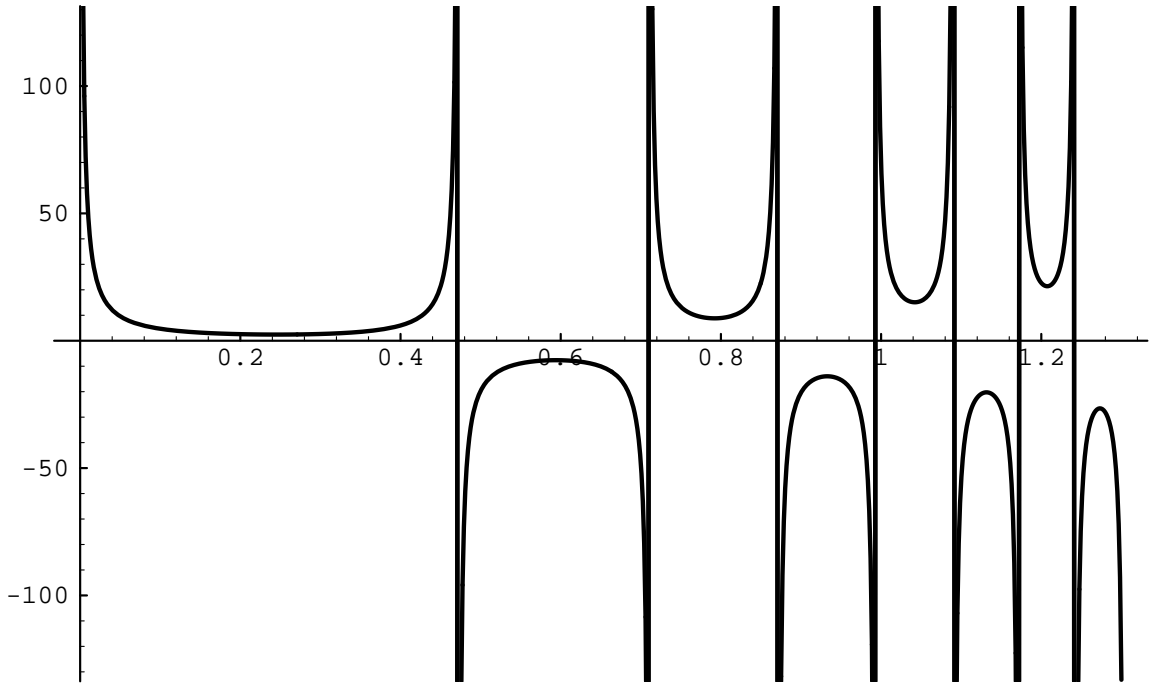}
	\caption{Plot of $\Omega^{'2}$ as a function of $t'$ for the
	oscillator of Eq.~(\ref{m-o}) with $m_0=\mu=\omega_0=1$.}
	\end{figure}
\section*{Acknowledgements}
I would like to thank Dr.~M.~Razavi and R.~Allahverdi for invaluable
discussions and acknowledge the financial support of the Killam Foundation
of Canada.

\end{document}